\input pipi.sty
\input epsf.sty
\magnification1020

\nopagenumbers
\rightline{FTUAM 03-15}
\rightline{September, 1, 2003}
\rightline{hep-ph/0309039}
\bigskip
\hrule height .3mm
\vskip.6cm
\centerline{{\bigfib The  
quadratic scalar radius of the pion and the mixed $\pi-K$ radius}\footnote{*}{
Research supported in part by CICYT, Spain.}}
\medskip
\centerrule{.7cm}
\vskip1cm

\setbox9=\vbox{\hsize65mm {\noindent\fib F. J. 
Yndur\'ain} 
\vskip .1cm
\noindent{\addressfont Departamento de F\'{\i}sica Te\'orica, C-XI\hb
 Universidad Aut\'onoma de Madrid,\hb
 Canto Blanco,\hb
E-28049, Madrid, Spain.}\hb}
\smallskip
\centerline{\box9}
\bigskip
\setbox0=\vbox{\abstracttype{Abstract} We consider the 
quadratic scalar radius of the pion, $\langle r^2_{{\rm S},\pi}\rangle$, and the 
mixed $K-\pi$ scalar radius, $\langle r^2_{{\rm S},K\pi}\rangle$. 
With respect to the second, we point out that the more recent 
(post-1974) experimental results in 
$K_{l3}$ decays imply a value, $\langle r^2_{{\rm S},K\pi}\rangle=0.31\pm0.06\,{\rm fm}^2$, 
which is about $2\,\sigma$ above estimates based on chiral 
perturbation theory. 
On the other hand, we show that this value of $\langle r^2_{{\rm S},K\pi}\rangle$
  suggests the existence 
of a low mass S$\tfrac{1}{2}$ $K\pi$ resonance.
With respect to $\langle r^2_{{\rm S},\pi}\rangle$, we contest the central value and accuracy of 
current evaluations, that give  
 $\langle r^2_{{\rm S},\pi}\rangle=0.61\pm0.04\,{\rm fm}^2$.  
Based on experiment, we find a robust lower bound
 of $\langle r^2_{{\rm S},\pi}\rangle\geq0.70\pm0.06\,{\rm fm}^2$ 
and a reliable estimate,  $\langle r^2_{{\rm S},\pi}\rangle=0.75\pm0.07\,{\rm fm}^2$, 
where the error bars are attainable. 
This implies, in particular, that the chiral 
result for $\langle r^2_{{\rm S},\pi}\rangle$ is  $1.4\,\sigma$ away from experiment.
We also comment on implications about the chiral parameter $\bar{l}_4$, 
very likely substantially larger (and with larger errors) than usually assumed. 
}
\centerline{\box0}
\brochureendcover{Typeset with \physmatex}
\brochureb{\smallsc f. j.  yndur\'ain}{\smallsc the  
quadratic scalar radius of the pion and the mixed {\petit $\pi-K$} radius }{1}

\booksection{1. Introduction}

\noindent
The quadratic scalar radius of the pion, $\langle r^2_{{\rm S},\pi}\rangle$, 
and the mixed $K-\pi$ (quadratic) scalar radius, 
$\langle r^2_{{\rm S},K\pi}\rangle$, are quantities of high interest for 
chiral perturbation theory calculations, or, more 
generally, for pion physics. 
Using chiral perturbation theory to one loop they can be related to 
meson masses and decay constants:\ref{1,2}
$$\eqalign{
\langle r^2_{{\rm S},K\pi}\rangle=&\,\dfrac{6}{M^2_K-M^2_\pi}
\left(\dfrac{f_K}{f_\pi}-1\right)+\delta_2;\cr
\delta_2=&\,-\dfrac{1}{192\pi^2f^2_\pi}\left\{
15h_2(M^2_\pi/M^2_K)+\dfrac{19 M_K^2+3M^2_\eta}{M^2_K+M^2_\eta}
h_2(M^2_\eta/M_K^2)-18\right\},\cr
h_2(x)=&\,\tfrac{3}{2}\left(\dfrac{1+x}{1-x}\right)^2+
\dfrac{3x(1+x)}{(1-x)^3}\log x;
\cr}
\equn{(1.1a)}$$
$$\eqalign{
\langle r^2_{{\rm S},\pi}\rangle=&\,\dfrac{6}{M^2_K-M^2_\pi}
\left(\dfrac{f_K}{f_\pi}-1\right)+\delta_3;\cr
\delta_3=&\,-\dfrac{1}{64\pi^2f^2_\pi}\,\dfrac{1}{M_K^2-M^2_\pi}\Bigg\{
6(2M_K^2-M^2_\pi)\log\dfrac{M_K^2}{M^2_\pi}+
9M^2_\eta\log\dfrac{M^2_\eta}{M^2_\pi}
-
2(M_K^2-M^2_\pi)\left(10+\tfrac{1}{3}\dfrac{M^2_\pi}{M^2_\eta}\right)\Bigg\}.
\cr}
\equn{(1.1b)}$$
The second can also be expressed in terms of the  chiral constant 
$\bar{l}_4$; to  one loop,\ref{1}
$$\langle r^2_{{\rm S},\pi}\rangle=\dfrac{3}{8\pi^2f^2_\pi}\Big\{\bar{l}_4-\tfrac{13}{12}\Big\}.
\equn{(1.2)}$$
Here $f_\pi,\,f_K$ and $M_\pi$, $M_K$ are the decay constants and masses of pion and kaon; 
$M_\eta=547\,\mev$ is the eta particle mass. 
We  take $M_\pi=139.57\,\mev$  (the charged pion mass), but choose an average kaon mass,
$M_K=496\,\mev$. 
From (1.1a,b), Gasser and Leutwyler\ref{2} obtain the theoretical predictions
$$\langle r^2_{{\rm S},K\pi}\rangle_{\rm GL}=0.20\pm0.05\;{\rm fm}^2,
\equn{(1.3a)}$$
$$\langle r^2_{{\rm S},\pi}\rangle_{\rm GL}=0.55\pm0.15\;{\rm fm}^2;
\equn{(1.3b)}$$
 the errors come from estimated higher order corrections.

For $\langle r^2_{{\rm S},K\pi}\rangle$ we have experimental information from the decays 
$K^0_{l3}$ and  $K^\pm_{l3}$. 
For the first the world average of the Particle Data Tables\ref{3} is
$$\lambda_0=0.025\pm0.006\quad [K^0_{l3}]
\equn{(1.4a)}$$
and $\langle r^2_{{\rm S},K\pi}\rangle=6\lambda_0/M^2_\pi$. For $K^\pm_{l3}$ the four more modern 
experimental analyses\ref{4} give the numbers\fnote{This is one of the few cases in which 
the PDT recommend a number difficult to believe. Perhaps influenced by very old 
determinations (pre-1975) they 
give the average value $0.006\pm0.007$, incompatible both with isospin 
invariance and with the with post-1974 experiments,  
and which we disregard.} 
$$\lambda_0=\cases{
0.062\pm0.024\qquad\hbox{Artemov et al.\quad (1997)}\cr
0.029\pm0.011\qquad\hbox{Whitman et al.\quad(1980)}\cr
0.019\pm0.010\qquad\hbox{Heintze et al.\quad(1977)}\cr
0.008\pm0.097\qquad\hbox{Braun et al.\quad(1975).}\cr
}
\equn{(1.4b)}$$
If we average them, which is permissible since they are compatible within errors, we find 
$$\lambda_0=0.027\pm0.007\quad [K^{\pm}_{l3}],
\equn{(1.4c)}$$
a value in perfect agreement with (1.4a), to which it should equal 
if neglecting isospin breaking effects. 
We compose (1.4a), (1.4c) to get $\lambda_0=0.026\pm0.005$, 
 and find  what we will consider the experimental value for 
the form factor:
$$\langle r^2_{{\rm S},K\pi}\rangle_{\rm exp.}=0.312\pm0.070\;{\rm fm}^2.
\equn{(1.5)}$$
On comparing with a dispersive calculation, (1.5) strongly suggests 
the existence of a low energy S$\tfrac{1}{2}$ $K\pi$ resonance. On the other hand, 
the central value in (1.5) lies clearly outside the error
 bars of the chiral theory prediction, (1.3a).
\goodbreak

There is no direct measurement of $\langle r^2_{{\rm S},\pi}\rangle$. 
 Donoghue, Gasser ad Leutwyler\ref{5} used 
the two-channel Omn\`es--Muskhelishvili method and $\pi\pi$  phase shifts to 
give what is presented as a  precise experiment-based estimate; Colangelo, 
Gasser and Leutwyler review it and, with a minor updating, accept it at present:\ref{6}
$$\langle r^2_{{\rm S},\pi}\rangle=0.61\pm0.04\;{\rm fm}^2.
\equn{(1.6)}$$
It is difficult to believe that 
the precision and central value in (1.6) hold at the same time.
 To get these numbers, Donoghue et al. use experimental phase shifts 
for $\pi\pi$ scattering above the $\bar{K}K$ threshold, 
where, because one does not measure the process  $\bar{K}K\to \bar{K}K$, 
the set of measurements is {\sl incomplete} (as proved for example in 
refs.~7) and where, indeed, different fits give totally different 
eigenphases  (necessary to perform the  Omn\`es--Muskhelishvili 
analysis), as may be seen explicitly in ref.~8. 
Moreover, they neglect multipion contributions
 which, for the electromagnetic form factor of the pion, 
account for some 6\% of the full result.

As a matter of fact, we will give here a new evaluation 
(which is the main outcome of the present note)  
and will, in particular, present  examples of 
phases which are compatible with experimental information, as well as
 with all physical requirements 
at high energy, 
and for which the corresponding $\langle r^2_{{\rm S},\pi}\rangle$ is 
several standard deviations above (1.6).
In particular, we find a safe bound, and a  reliable estimate:
$$\langle r^2_{{\rm S},\pi}\rangle\geq0.70\pm0.06\;{\rm fm}^2,
\equn{(1.7)}$$
$$\langle r^2_{{\rm S},\pi}\rangle=0.75\pm0.07\;{\rm fm}^2.
\equn{(1.8)}$$
Moreover, we show that the error bars in (1.8) are {\sl attainable}.

It thus follows that, also for $\langle r^2_{{\rm S},\pi}\rangle$, 
 the errors due to higher orders are underestimated.

\booksection{2. The Omn\`es--Muskhelishvili method for form factors and radii}

\noindent 
We consider the scalar pion form factor, $F_S(t)$, and the mixed scalar form factor 
$f_{K\pi}(t)$. We will also discuss the electromagnetic form factor of the pion,
$F_\pi(t)$. In terms of these,\fnote{We define the form factors by
$$\eqalign{
\langle\pi(p)|J_{\rm e.m.}^\mu(0)|\pi(p')\rangle=&\,(2\pi)^{-3}(p^\mu-p'^{\mu})F_\pi(t),\cr
\langle\pi(p)|m_u\bar{u}u(0)+m_d\bar{d}d(0)|\pi(p')\rangle=&\,(2\pi)^{-3}F_S(t),\cr
\langle\pi(p)|(m_s-m_u)\bar{u}s(0)|K(p')\rangle=&\,(2\pi)^{-3}f_{K\pi}(t);
\cr}$$
the meson states are normalized to $\langle p|p'\rangle=
2p_0\delta({\bf p}-{\bf p}')$, and $t=(p-p')^2$.}
$$\eqalign{
F_\pi(t)\simeqsub_{t\to0}&\,F_\pi(0)\Big\{1+\tfrac{1}{6}\langle r^2_\pi\rangle\,t\Big\}, \cr
F_S(t)\simeqsub_{t\to0}&\,F_S(0)\Big\{1+\tfrac{1}{6}\langle r^2_{{\rm S},\pi}\rangle\,t\Big\}, \cr
f_{K\pi}(t)\simeqsub_{t\to0}&\,f_{K\pi}(0)
\Big\{1+\tfrac{1}{6}\langle r^2_{{\rm S},K\pi}\rangle\,t\Big\}, \cr
}
\equn{(2.1)}$$
and $\langle r^2_\pi\rangle$ is the electromagnetic radius of the pion. 
For $F_\pi$, current conservation gives $F_\pi(0)=1$.
The values of $F_S(0)$, $f_{K\pi}(0)$ may be calculated with 
chiral dynamics,\ref{2}
but we will not concern ourselves with this here.

Let us denote by $F(t)$ to any of the three form factors in (2.1), and let 
$\delta(t)$ be its phase:
$$\delta(t)=\arg F(t),\quad t\geq s_{\rm th};
\equn{(2.2)}$$
$s_{\rm th}$ is the threshold, $4M^2_\pi$ or $(M_\pi+M_K)^2$, as the case may be. 
(We note that in (2.2) we do {\sl not} understand the  principal value of the argument; 
the phase has to be taken as varying continuously with $t$).
The Fermi--Watson final state interaction theorem implies that, 
for $t<s_0$ (where $s_0$ is the energy at which 
inelastic channels become nonnegligible), $\delta(t)$
 equals a corresponding scattering phase.\fnote{
Actually, the individual relations in (2.3) also hold at any $t$ (even above $s_0$) provided 
inelasticity for the corresponding partial wave is negligible there.} 
To be precise,
$$\eqalign{
\delta(t)=\delta_1(t)\quad\hbox{[P wave $\pi\pi$ phase, for $F_\pi$]};
\quad s_0\simeq 1.1\;{\gev}^2\cr
\delta(t)=\delta_0^{(0)}(t)\quad\hbox{[S0 wave (S wave with isospin 0) $\pi\pi$ phase, for $F_S$]};
\quad s_0=4M_K^2\cr
\delta(t)=\delta_0^{(1/2)}(t)\quad
\hbox{[S$\tfrac{1}{2}$ (S wave with isospin $\tfrac{1}{2}$)  $K\pi$ phase, for $f_{K\pi}$]};
\quad s_0\simeq 1.5^2\;{\gev}^2.\cr
}
\equn{(2.3)}$$
We will assume that we know the phases $\delta_1$, $\delta_0^{(0)}$, $\delta_0^{(1/2)}$, 
and thus $\delta(t)$, 
for $t\leq s_0$.

At large $t$,  the Brodsky--Farrar counting rules\ref{9} imply  that
$$F(t)\simeqsub_{t\to\infty}\dfrac{1}{-t\log^\nu (-t)},
\equn{(2.4)}$$
from which it follows that, unless the phase oscillated at infinity, one must have
$$\delta(t)\simeqsub_{t\to+\infty}
\pi\left\{1+\nu\dfrac{1}{\log t/\hat{t}}\right\}.
\equn{(2.5)}$$
In particular, (2.5) implies that $\delta(\infty)=\pi$.
For $F_\pi$, the Jackson--Farrar calculation\ref{9} gives
$$F_\pi(t)\simeqsub_{t\to\infty}12\pi C_Ff^2_\pi\alpha_s(-t)/(-t),$$
hence $\nu=1$; 
 for the other form factors one cannot  
 prove a similar behavior rigorously in QCD, 
although it is likely that $\nu=1$ also here. 
$\hat{t}$ is a scale; for the electromagnetic form factor, it is  
$\sim\lambdav^2$, with $\lambdav$ the QCD parameter, but its precise value is 
generally not known.
Nevertheless, the feature that the limit $\delta(\infty)$ 
has to be reached {\sl from above}, i.e., that at 
asymptotic energies  $\delta(t)$ is {\sl larger} than $\pi$, seems to be general.

We will use the Omn\`es--Muskhelishvili method,\ref{10} with only one channel, to solve 
for  $F$ in terms of $\delta$; it will turn out 
that the two-channel method is neither necessary nor reliable 
[the last for the reasons explained  after \equn{(1.6)}]. According to it, we have that,
given the condition (2.5), the phase determines uniquely $F$: one has 
$$F(t)=F(0)\exp\left\{\dfrac{t}{\pi}
\int_{s_{\rm th}}^\infty\dd s\,\dfrac{\delta(s)}{s(s-t)}\right\}.
\equn{(2.6)}$$
From this we get a simple sum rule for the square radius 
$\langle r^2\rangle$ corresponding to $F(t)$:
$$\langle r^2\rangle=\dfrac{6}{\pi}\int_{s_{\rm th}}^\infty\dd s\,\dfrac{\delta(s)}{s^2}.
\equn{(2.7)}$$
In general, we will split $\langle r^2\rangle$ as follows:
$$\langle r^2\rangle=Q_J(s_0)+Q_\phiv(s_0)+Q_G(s_0).
\equn{(2.8)}$$
Here $Q_J$ is the piece in (2.7) coming from the region 
where we know $\delta$,
$$Q_J(s_0)\equiv\dfrac{6}{\pi}\int_{s_{\rm th}}^{s_0}\dd s\,\dfrac{\delta(s)}{s^2}.
\equn{(2.9a)}$$
$Q_\phiv$ is obtained defining an effective phase 
that interpolates linearly (in $t^{-1}$) between the values 
of $\delta(t)$ at $s_0$ and $\infty$: we write
$$\delta_{\rm eff}(t)\equiv\pi+\Big[\delta(s_0)-\pi\Big]\dfrac{s_0}{t},
\equn{(2.9b)}$$
and then set
$$Q_\phiv(s_0)\equiv\dfrac{6}{\pi}\int_{s_0}^{\infty}\dd s\,\dfrac{\delta_{\rm eff}(s)}{s^2}.
\equn{(2.9c)}$$
Finally, $Q_G$ corrects for the difference between $\delta$ and $\delta_{\rm eff}$:
$$Q_G(s_0)\equiv\dfrac{6}{\pi}\int_{s_0}^{\infty}\dd s\,\dfrac{\delta(s)-\delta_{\rm eff}(s)}{s^2}.
\equn{(2.9d)}$$
$Q_J$, $Q_\phiv$ are known; $Q_G$ has to be fitted or estimated. 
The decomposition (2.8) is equivalent to decomposing $F$ as a product. 
We integrate explicitly $\delta_{\rm eff}$ and then we can write
$$\eqalign{
F(t)=&\,F(0)J(t)\phiv(t) G(t);\cr
J(t)=&\,\exp\left\{\dfrac{t}{\pi}\int_{s_{\rm th}}^{s_0} \dd s\;
\dfrac{\delta(s)}{s(s-t)}\right\},\cr
\phiv(t)=&\,\ee^{1-\delta_1^1(s_0)/\pi}
\left(1-\dfrac{t}{s_0}\right)^{[1-\delta_1^1(s_0)/\pi]s_0/t}
\left(1-\dfrac{t}{s_0}\right)^{-1};\cr
G(t)=&\,\exp\left\{\dfrac{t}{\pi}\int_{s_0}^\infty \dd s\;
\dfrac{\delta(s)-\delta_{\rm eff}(s)}{s(s-t)}\right\}.\cr
}
\equn{(2.10)}$$
  
What we know about $G(t)$ is that $G(0)=1$, and that it 
 is analytic except for the cut $s_0\leq t<\infty$.
 The best way to take this into account is by making a conformal mapping 
of this cut plane into a disk, and expand in the conformal variable, $z(t)$:  
$$z(t)=\dfrac{\tfrac{1}{2}\sqrt{s_0}-\sqrt{s_0-t}}{\tfrac{1}{2}\sqrt{s_0}+\sqrt{s_0-t}}.
\equn{(2.11a)}$$
We then write\ref{11}
$$G(t)=1+A_0+c_1z+c_2 z^2+c_3 z^3+\cdots\,,
\equn{(2.11b)}$$
an expansion that will be convergent for all $t$ inside the cut plane. 
We can implement the condition $G(0)=1$, order by order, by writing
$A_0=-\left[c_1z_0+c_2 z_0^2+c_3 z_0^3+\cdots\right]$,
$z_0\equiv z(t=0)=-1/3;$
the expansion then reads,

$$G(t)=1+c_1(z+1/3)+c_2 (z^2-1/9)+c_3 (z^3+1/27)+\cdots,
\eqno{(2.12)}$$
the $c_i$ being free parameters.

The contributions to the square radius $Q_\phiv$, $Q_G$ may be written 
explicitly in terms of $\delta(s_0)$, 
$c_i$ as
$$Q_\phiv=\dfrac{3}{s_0}\left\{1+\dfrac{\delta(s_0)}{\pi}\right\},\qquad
Q_G=\dfrac{4}{3s_0}\left\{c_1-\tfrac{2}{3}c_2+\cdots\right\}.
\eqno{(2.13)}$$
For the electromagnetic form factor of the pion 
we  take, following ref.~11, $s_0=1.1\,\gev^2$. 
For $F_\pi$ we can fit experimental data 
and thus find the $c_i$. 
These data are in fact precise enough to give two terms:\ref{11}
$$c_1=0.38\pm0.03,\quad c_2=-0.19\pm0.03\qquad\hbox{[For $F_\pi$].}
\eqno{(2.14)}$$
The squared charge radius is then
$$\langle r^2_\pi\rangle=0.435\pm0.003\;{\rm fm}^2.
\eqno{(2.15)}$$

\booksection{3. Dispersive evaluation of the square radii}
\vskip-0.5truecm
\booksubsection{3.1. The electromagnetic radius of the pion}

\noindent
We start with a review of the evaluation of $\langle r^2_\pi\rangle$, which will 
serve as a model for the other two. 
Although in ref.~11 the P wave phase was {\sl deduced} from the 
experimental values of $F_\pi$, we here consider it as given, 
for $4M^2_\pi\leq t\leq 1\,{\gev}^2$. 
Taking its value from ref.~11   
 one has (in ${\rm fm}^2$)
$$Q_J=0.195.
\eqno{(3.1a)}$$
Moreover,  $\delta(s_0)=\delta_1(s_0)=2.70$, and hence
$$Q_\phiv=0.217.
\equn{(3.1b)}$$
So, if we approximated $G(t)\equiv 1$ (physically, 
this is approximately equivalent to neglecting inelastic channels), 
 we would underestimate the radius, but not by much, as we get
$$Q_J+ Q_\phiv=0.412,$$
which is 6\% below the full value of $\langle r^2_\pi\rangle$ as given in (2.15).

It is not easy to guess $Q_G$, although it is easy to understand its sign: 
in (2.5) we have interpolated linearly
 from $s_0=1\,\gev^2$, where $\delta(s_0)<\pi$, to $\delta(\infty)=\pi$; 
that is to say, systematically {\sl below} the value $\pi$, 
while we know, from (2.5), that $\delta(t)$ must approach the asymptotic value of 
$\pi$ {\sl from above}. 
Thus  the phase $\delta(s)$ should rise beyond $\pi$, 
probably around the energy of the resonance 
$\rho(1450)$, to reach its asymptotic behaviour (above $\pi$) 
after that. 
So we expect 
$$Q_G\sim6\int_{s_{\rm as.}}^\infty\dd s\, \dfrac{1}{\log s/\hat{t}},
\quad
s^{1/2}_{\rm as.}\gsim 1.45\;\gev.
\equn{(3.2)}$$
Indeed, using this formula with reasonable choices of $s_{\rm as.}$,
 $\hat{t}$, we get a value near the 
experimentally measured one for $Q_G$; for example, we obtain the exact result, $Q_G=0.024$, 
with $\hat{t}\simeq0.3\,\gev^2$,
  $s_{\rm as.}^{1/2}\simeq1.8\,\gev$.

\booksubsection{3.2. The mixed $K\pi$ scalar radius}

\noindent
We  first assume the phase $\delta_0^{(1/2)}(t)$ to be given, for $t^{1/2}\leq 1.5\,\gev$, 
by the resonance $K^*(1430)$, whose properties we take from the PDT.\ref{3} 
Its mass is $M_*=1412\pm6 \mev$, and its width $\gammav_*=294\pm23\,\mev$; 
we neglect its small inelasticity ($\sim7\%$). 
We write a Breit--Wigner formula for the phase:

$$\cot\delta_0^{(1/2)}(t)=\dfrac{t^{1/2}}{2q}(1-s/M^2_*)B_0,\quad
 q=\dfrac{\sqrt{[s-(M_K-M_\pi)^2][s-(M_K+M_\pi)^2]}}{2s^{1/2}}
$$
and $B_0=2q(M^2_*)/\gammav_*=4.15\pm0.35.$
We take $s_0=1.5^2\,\gev^2$, and then we
 have
$$
Q_J=0.050\pm0.025,\quad Q_\phiv=0.087\pm0.001;\qquad
Q_J+Q_\phiv=0.137\pm0.03.
\equn{(3.3a)}$$
This means that $Q_G$ is large; in fact, 
on comparing with the experimental value, \equn{(1.5)}, 
we find
$$Q_G=0.175\pm0.03.
\equn{(3.3b)}$$
The corresponding $c_1$ would also be large, $c_1=7.6$

The sum of $Q_J$ and $Q_\phiv$ substantially {\sl underestimates} 
the value of the mixed scalar square radius: 
the true phase $\delta(t)$ of the form factor would have to
 go on growing 
a lot before  
setting to the asymptotic regime (2.5). 
 The size of the phase necessary 
to produce the large $Q_G$ required appears excessive. 

An alternate possibility is the existence of a lower  energy resonance (or enhancement; 
we denote it by $\kappa$),  
{\sl below} the $K^*(1430)$, which some analyses suggest,\ref{12}
with $M_\kappa\sim 1\,\gev$ and $\gammav_\kappa=400\pm100\,\mev$. 
In this case, we approximate the low energy phase, $s\leq s_0=1\,\gev^2$, by writing\fnote{The 
Breit--Wigner parametrization (3.4) for the $\kappa$ should be considered 
only as an effective one; it is in fact not clear that the 
 phase would 
reach $90\degrees$ at $M_\kappa$.}
$$\cot\delta_0^{(1/2)}(t)=\dfrac{t^{1/2}}{2q}(1-s/M^2_\kappa)B_\kappa,\quad
B_\kappa=1.8\pm0.5
$$
and find
$$
Q_J=0.070\pm0.030,\quad Q_\phiv=0.180\pm0.006;\qquad
Q_J+Q_\phiv=0.250\pm0.030,
\equn{(3.4a)}$$
which reproduces well the experimental number with a small $Q_G$, compatible with zero: 
$$Q_G\sim 0.06\pm0.07.
\equn{(3.4c)}$$

\booksubsection{3.3. The scalar radius of the pion: bounding its value}

\noindent
Next we consider the quadratic scalar radius of the pion.
We will, for the S0 phase below $t^{1/2}=0.96\,\gev$, take the two fits 
to experimental data in 
ref.~13: one possibility is  
$$\eqalign{
\cot\delta_0^{(0)}(s)=&\,\dfrac{s^{1/2}}{2k}\,\dfrac{M_{\pi}^2}{s-\tfrac{1}{2}M_{\pi}^2}\,
\dfrac{M^2_\sigma-s}{M^2_\sigma}\,
\left\{B_0+B_1\dfrac{\sqrt{s}-\sqrt{4M_K^2-s}}{\sqrt{s}+\sqrt{4M_K^2-s}}\right\},
\quad  k=\sqrt{s/4-M^2_\pi};\cr
{B}_0=&\,21.04,\quad {B}_1=6.62,\quad
M_\sigma=782\pm24\,\mev;\cr
{\chi}^2/{\rm d.o.f.}=&\,15.7/(19-3);\quad 
a_0^{(0)}=(0.230\pm0.010)\,M^{-1}_\pi.
\cr    }
\equn{(3.5a)}$$
Uncorrelated errors are obtained if 
replacing the $B_i$ by the 
parameters $x,\,y$ with
$$B_0=y-x;\quad B_1=6.62-2.59 x:\quad y=21.04\pm0.75,\quad x=0\pm 2.4.
\equn{(3.5b)}$$
This will be referred to as 2Bs. Alternatively, we may take 
$$\eqalign{
\cot\delta_0^{(0)}(s)=&\,\dfrac{s^{1/2}}{2k}\,\dfrac{M_{\pi}^2}{s-\tfrac{1}{2}M_{\pi}^2}\,
\dfrac{M^2_\sigma-s}{M^2_\sigma}\,
\left\{B_0+B_1\dfrac{\sqrt{s}-\sqrt{4M_K^2-s}}{\sqrt{s}+\sqrt{4M_K^2-s}}+
B_2\left[\dfrac{\sqrt{s}-\sqrt{4M_K^2-s}}{\sqrt{s}+\sqrt{4M_K^2-s}}\right]^2\right\};\cr
\quad\chi^2/{\rm d.o.f.}=&\,11.1/(19-4);\quad
 a_0^{(0)}=(0.226\pm0.015)\;M_{\pi}^{-1}\cr
\quad M_\sigma=806\pm21,&\,\quad B_0=21.91\pm0.62,\quad B_1=20.29\pm1.55, \quad B_2=22.53\pm3.48;\cr
}
\equn{(3.6)}$$
this we denote by 3Bs. 

Although we think 2Bs to be more close to reality than 3Bs, and although 
both give very similar results, we include 3Bs because it comprises, within 
its errors, the S0 phase shift by Colangelo, Gasser and Leutwyler,\ref{6} 
which these authors present as very precise and incorporating 
results from chiral dynamics (in addition to 
analyticity and unitarity). 
We find, with self-explanatory notation,
$$Q_J(t^{1/2}\leq0.96\,\gev; {\rm 2Bs})=0.452\pm0.05\;{\rm fm}^2,
\equn{(3.7a)}$$
$$Q_J(t^{1/2}\leq0.96\,\gev; {\rm 3Bs})=0.440\pm0.05\;{\rm fm}^2.
\equn{(3.7b)}$$

Between $t^{1/2}=0.96\,\gev$ and $\bar{K}K$ threshold, that we take 
at $t^{1/2}=0.992\,\gev$, we use a fit to experimental data as given in 
\equn{(3.8)} of ref.~13(a), 
 and get, in ${\rm fm}^2$,
$$Q_J(0.96\,\gev\leq t^{1/2}\leq0.992\,\gev)=0.013\pm0.002.
$$
We then find the numbers 
$$Q_J=\cases{
0.465\pm0.05\quad[{\rm 2Bs}]\cr
0.453\pm0.05\quad[{\rm 3Bs}].\cr}
\equn{(3.8)}$$

To calculate $Q_\phiv$ we take the value
$$\delta_0^{(0)}(4M_K^2)=3.14\pm0.52,$$
which covers all the experimental determinations,\ref{14}
and get
$Q_\phiv=0.237\pm0.02.$
Therefore, we have obtained the result
$$Q_J+Q_\phiv=0.70\pm0.06,
\equn{(3.9)}$$
and this  comprises both cases 2Bs and 3Bs.

Eq.~(3.9) should be interpreted as providing a {\sl lower bound}
 on $\langle r^2_{{\rm S},\pi}\rangle$;
 it assumes that the 
phase of $F_S(s)$ does not increase 
for $s$ beyond $\bar{K}K$ threshold, while, as one would 
deduce from the similar calculation of $\langle r^2_\pi\rangle$,
and as we will see also in the present case,  
 $\delta(s)$ should {\sl increase} somewhat before decreasing to its asymptotic value, 
$\delta(\infty)=\pi$. 
We have therefore found the result,
$$\langle r^2_{{\rm S},\pi}\rangle\geq 0.70\pm0.06\;{\rm fm}^2.
\equn{(3.10)}$$

\booksubsection{3.4. The scalar radius of the pion:  calculations}

\noindent
We can get a first {\sl estimate} of the remaining quantity needed to 
calculate $\langle r^2_{{\rm S},\pi}\rangle$, $Q_G$, by 
invoking SU(3) invariance. 
If the $\kappa$ is the SU(3) partner of the $\sigma$, 
we inded expect $M_\kappa\simeq1\,\gev$. 
Identifying $Q_G(K\pi)\simeq Q_G(\pi)$, and using (3.4), we find  an
approximate number, 
$$\langle r^2_{{\rm S},\pi}\rangle\simeq0.72\pm0.09\;{\rm fm}^2.
\equn{(3.11)}$$

A more sophisticated method to get  $Q_G$ is as follows. 
As implied by the experimental data on $\pi\pi$ scattering,\ref{14({\rm b})} 
the inelasticity is compatible with zero (indeed, the central value is 
almost equal to zero) for the S0 wave, within experimental 
errors, in the 
energy region $1.1\,\gev\leq s^{1/2}\leq 1.5\,\gev$. 
It thus follows that the phase of $F_S(s)$ 
must  be approximately equal to $\delta_0^{(0)}(s)$   
for $1.1\,\gev\leq t^{1/2}\leq1.42\,\gev$.

The phases $\delta_0^{(0)}(s)$, $\delta(s)$ will likely not be equal 
between $0.992\,\gev$ and $1.1\,\gev$; however, 
because this is a very short range, and the 
phases are  are equal at both endpoints (in the 
approximation of neglecting inelasticity there), it follows that any 
reasonable interpolation, e.g., a linear interpolation, will give results not 
 very different from what one gets by taking, simply,
$$\delta(s)=\delta_0^{(0)}(s),\; \hbox{in the full  region,}\;
  0.992\,\gev\leq s^{1/2}\leq1.42\,\gev.$$  
The distortion  caused by the inelasticity being nonzero
 just around $1\,\gev$
 is negligible, numerically;
 later we will add the estimated error due to above relation being only approximately
true.

\topinsert{
\setbox0=\vbox{\hsize15.9truecm{\epsfxsize 14.truecm\epsfbox{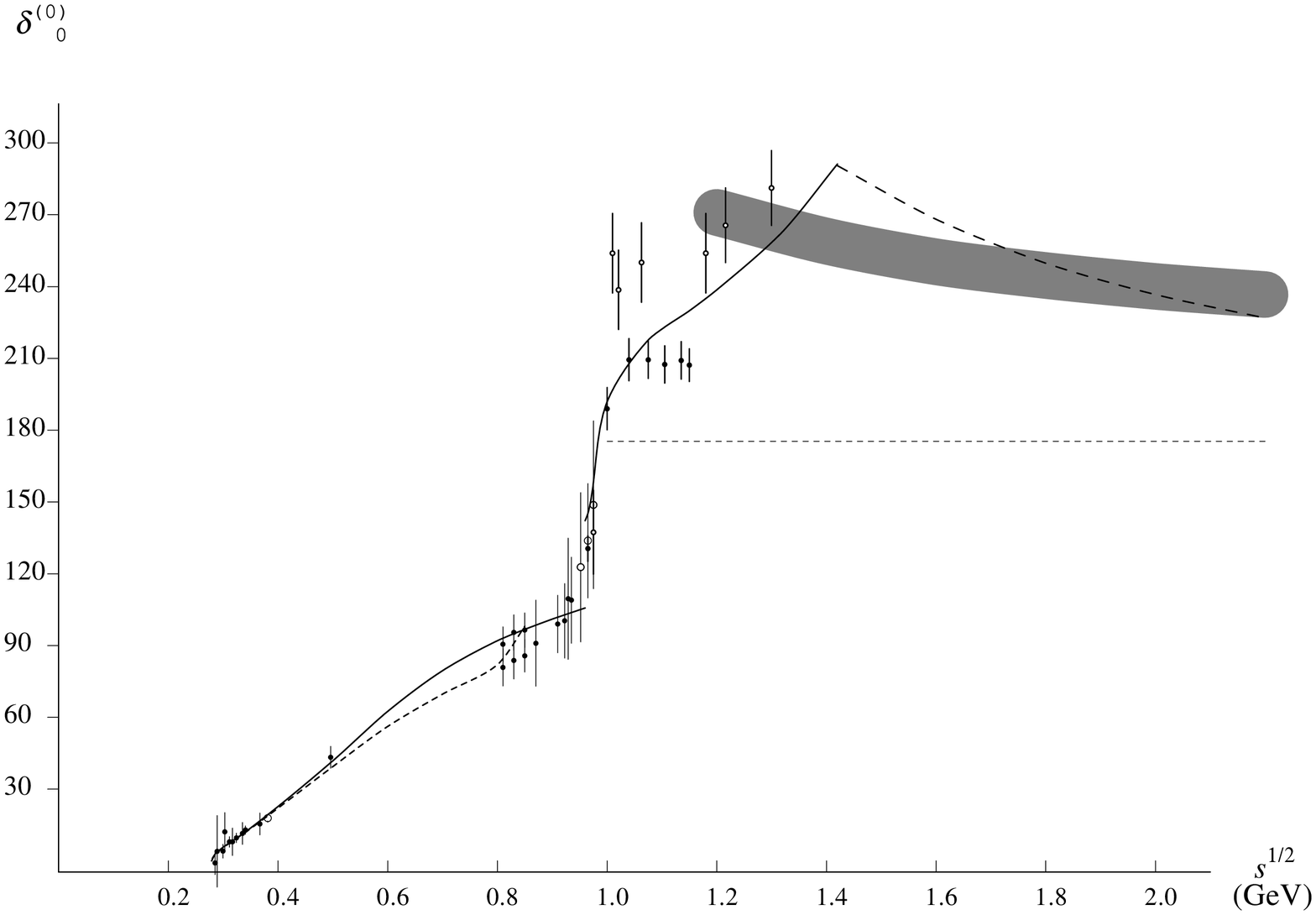}}} 
\setbox6=\vbox{\hsize 16truecm\captiontype\figurasc{Figure 1. }{
The  
$I=0$, $S$-wave phase shift corresponding to 2Bs and Eq.~(3.12) (continuous line) 
and the phase of ref.~6 (dotted line at low energy). 
Also shown  experimental data points 
  included in the 2Bs fit:  from the $K_{l4}$ experiments,
and a point from $K_{2\pi}$ decay; 
 high energy data of Protopopescu et al. (black dots), and Hyams et al., 
Grayer et al. (open circles), and the $s$-channel solution 
of  Estabrooks and Martin. 
The dotted line at high energy is the line $\delta=\pi$; the 
dashed line is the interpolation corresponding to $\delta_{\rm eff}$, Eq.~(2.9b). 
The asymptotic phase $\delta_{\rm as.}$, with   $0.1\,{\rm GeV}^2\leq \hat{t}\leq0.35\,{\rm GeV}^2$,
 is represented 
by the thick gray line.}\hb} 
\centerline{\tightboxit{\box0}}
\bigskip
\centerline{\box6}
}\endinsert 

We take for $\delta_0^{(0)}(s)$ the fit to experimental data in ref.~13(a), 
\equn{(3.8)}, 
$$\eqalign{
\cot\delta_0^{(0)}(s)=&\,c_0\dfrac{(s-M^2_\sigma)(M_f^2-s)|k_2|}{M^2_fs^{1/2}k_2^2};\quad
 k_2=\sqrt{s/4-M^2_K};\cr
s^{1/2}\geq0.96\;\gev;\quad c_0=&\,1.36\pm0.05,\quad
 M_\sigma=0.802\,\gev,\;M_f=1.32\,\gev.
\cr}
\equn{(3.12)}$$
The corresponding $\delta_0^{(0)}$ is shown in \fig~1. 
We write, choosing the 2Bs fit for the S0 wave below 
$\bar{K}K$ threshold,
$$Q_J(s_0=1.42^2\,{\gev}^2)=Q_J(0.992^2\,{\gev}^2)+Q_J(0.992^2\;{\rm to}\;1.42^2\,{\gev}^2),$$
and we have, in ${\rm fm}^2$,
$$\eqalign{
Q_J(0.992^2\,{\gev}^2)=&\,0.465\pm0.05, \quad
Q_J(0.992^2\;{\rm to}\;1.42^2\,{\gev}^2)=0.162\pm0.002;\cr
&\,Q_J(s_0=1.42^2\,{\gev}^2)=0.627\pm0.05.\cr
}$$
We note that the error in $Q_J(0.99^2\;{\rm to}\;1.42^2\,{\gev}^2)$ is only the 
error coming from $c_0$ in (3.12); the error 
due to neglect of the inelasticity 
we expect to be much larger, of the order of  10\% to 15\%.

In our present approximation, neglecting inelasticity 
below 1.42 \gev, we 
 have
 $\delta(1.42^2\;{\gev}^2)=\delta_0^{(0)}(1.42^2\;{\gev}^2)=5.10\pm0.03$, hence
$$Q_\phiv(s_0=1.42^2\;{\gev}^2)=0.152\pm0.001.$$
Thus, in this  calculation, and adding the estimated error due to neglect 
of inelasticity between 1 \gev\ and 1.42 \gev, we find
$$\langle r^2_{{\rm S},\pi}\rangle= 
0.78\pm0.06\;{\rm (St.)\,}^{+0}_{-0.07}\;{\rm (Inelast.)}\;\;{\rm fm}^2,
\equn{(3.13)}$$
 compatible with  (3.11). 
Although the 
central value here is probably displaced upwards 
(after all, there  is {\sl some} inelasticity), 
so that (3.13) should probably be considered more like an 
{\sl upper} bound, we  emphasize that this value 
is {\sl attainable}. 
Because experimental data are, at rather less than $1\,\sigma$, compatible 
with zero inelasticity, 
it follows that any realistic 
estimate for $\langle r^2_{{\rm S},\pi}\rangle$ must have error bars containing the value 
(3.13). 
This is one of the reasons 
why a two-channel evaluation is superfluous.

It is suggestive that, if we take the asymptotic formula 
(2.5) for $\delta(t)$,  with $\hat{t}$ between $0.1\,\gev^2$ and $0.35\,\gev^2$, then this
 coincides, on the average and 
to a 10\% accuracy, with the  $\delta_0^{(0)}(t)$, $\delta_{\rm eff}(t)$
[the second as given by (3.12), (2.9b) with $s_0=1.42^2\,\gev^2$],
 for $t^{1/2}$ between $1.1\,\gev$ and $2\,\gev$; see Fig.~1. 
This lends additional credence to our calculation (3.13), 
and it also suggests a different method of evaluation. 
We note that the asymptotic expression,
$$\delta_{\rm as.}(s)=\pi\left\{1+
{{\nu}\over{\log s/\hat{t}}}\right\},
$$
with $\nu=1$,
intersects the phase given by Eq.~(3.12) at $s\simeq1.35^2\,{\rm GeV}^2$, 
for $0.1\,{\rm GeV}^2\leq \hat{t}\leq0.35\,{\rm GeV}^2$. 
We can then use (3.12) for $s\leq1.35^2\,{\rm GeV}^2$ and 
the asymptotic expression $\delta_{\rm as.}$, with
  $0.1\,{\rm GeV}^2\leq \hat{t}\leq0.35\,{\rm GeV}^2$, for  $s\geq1.35^2\,{\rm GeV}^2$. 
This gives  
$$\langle r^2_{{\rm S},\pi}\rangle=0.76\pm0.06
\;{\rm fm}^2,
\eqno{(3.14)}$$
i.e., a result almost identical to (3.13).
Indeed, any reasonable interpolation between the asymptotic phase and 
the low energy one would give a similar result.

Taking into account this, as well as the 
previous  results, we get what we consider a reliable value for 
the scalar radius by writing
$$\langle r^2_{{\rm S},\pi}\rangle= 0.75\pm0.06\;{\rm fm}^2,
\equn{(3.15)}$$
which encompasses (3.10), 
 (3.11), (3.13) and (3.14).

\booksection{4. Discussion}

\noindent
We first say a few words about $\langle r^2_{{\rm S},K\pi}\rangle$. 
The central experimental value, 
$\langle r^2_{{\rm S},K\pi}\rangle_{\rm exp.}=0.31\pm0.06\;{\rm fm}^2$, 
is $2\,\sigma$ above the theoretical prediction of 
Gasser and Leutwyler,\ref{2} $0.20\pm0.05\;{\rm fm}^2$. 
It would seem that the errors
  were underestimated  by a factor $\sim2$ by these authors. 
The experimental number, together with 
our dispersive evaluation,  suggest the existence of 
the $\kappa$ enhancement.\ref{12}

We next turn to the scalar radius. 
 The  experiment-based evaluation of Donoghue, Gasser and Leutwyler,\ref{5,6} 
$\langle r^2_{{\rm S},\pi}\rangle_{\rm DGL}=0.61\pm0.04\;{\rm fm}^2$,
lies {\sl below} our lower bound, (3.10), and well    
below our best estimate, $0.75\pm0.07\;{\rm fm}^2$. 
To get a value as low as that of these authors,  
one would have to assume that, 
contrary to indications from $\pi\pi$ scattering, the 
phase of $F_S(t)$ would {\sl decrease} above $t^{1/2}=1\,\gev$ to about \ffrac{1}{2} of its 
asymptotic value; 
and that this continues to hold
 up to a very high energy.\fnote{It is difficult to point out where lies the 
failure in the calculation of  Donoghue, Gasser and Leutwyler, as it is of 
the ``black-box" type. However, a hint is obtained 
from their statement (p.~356 of ref.~5) that the values of the phases above $s^{1/2}=1.4\,\gev$ 
do not significantly affect their results. 
Contrary to this, our explicit calculations show that 
the contributions to (2.7) from energies above 1.4 \gev\ are {\sl large}\/: of 
20\% for $\langle r^2_{{\rm S},\pi}\rangle$, and of $27\%$ for the electromagnetic 
radius, $\langle r^2_\pi\rangle$, where one can check the estimate against experiment. 
They also provide 28\% of $\langle r^2_{{\rm S},K\pi}\rangle$.} 
This is of course a very unlikely behavior and, what is worse 
for a calculation   based upon experimental phase shifts, it is 
incompatible with what one may get within experimental errors, 
as proved by our  calculation (3.13). From this it follows  
that the chiral dynamics calculation at one loop,\ref{2}
$\langle r^2_{{\rm S},\pi}\rangle=0.55\pm0.15\;{\rm fm}^2$, lies 
clearly below   
the  value suggested by experiment, \equn{(3.15)}.

To finish we say a few words on the value of the chiral constant $\bar{l}_4$, 
and the connection with $\pi\pi$ parameters. 
We present a few values for  $\bar{l}_4$, all of them, however, using the 
DGL\ref{5} value for $\langle r^2_{{\rm S},\pi}\rangle$:
$$\eqalign{
\bar{l}_4=&\,4.4\pm0.3\quad\hbox{[BCT]};\quad
\bar{l}_4=4.2\pm0.2\quad\hbox{[ABT]};\cr
\bar{l}_4=&\,4.2\pm1.0\quad\hbox{[DFGS]};\quad
\bar{l}_4=4.4\pm0.2\quad\hbox{[Best CGL value; with $\pi\pi$ info]}.\cr
\cr}
\equn{(4.1)}$$
Here  CGL is ref.~6, ABT and BCT are in ref.~15, 
and DFGS denotes the paper by Descotes  et al.\ref{16} 
This is to be compared to what one gets, at one loop accuracy, from our results here:
$$\eqalign{
\bar{l}_4=&\,5.4\pm0.5,\quad\hbox{with our best estimate, (3.15)};\cr
\bar{l}_4\geq&\,5.1\pm0.4,\quad\hbox{with our bound (3.10)}.
\cr}
\equn{(4.2)}$$
Two loop corrections to the various form factors have been evaluated 
in ref.~17; see also ref.~6.
If we accepted the value given by Colangelo, Gasser and Leutwyler\ref{6} for the 
higher order correction, $\delta \bar{l}_4\simeq -0.25$, we would still find that 
their estimate in (4.1) is too low. 
This should have implications for the accuracy of their 
description of $\pi\pi$ scattering, as already mentioned by Descotes et al.,\ref{16}
who get, from fits including realistic errors for the $\pi\pi$
 phase shifts, much larger errors for $\bar{l}_4$ than the rest.  

In what respects to the connection with low energy 
$\pi\pi$ scattering parameters, our value here for
 $\langle r^2_{{\rm S},\pi}\rangle$ is in reasonable agreement 
with the D wave scattering lengths deduced in ref.~13, using the 
Froissart--Gribov representation with correct Regge expressions at high energy. 
From the relation
$$1+\tfrac{1}{3}M^2_\pi\langle r^2_{{\rm S},\pi}\rangle=24\pi f^2_\pi\Big\{a_1
-\tfrac{10}{3}M^2_\pi\left(a_2^{(0)}-\tfrac{5}{2}a_2^{(2)}\right)\Big\}M_\pi
-\tfrac{19}{576}\,\dfrac{M^2_\pi}{\pi^2 f^2_\pi},
$$
and using the numbers\fnote{Specifically, we take
 $a_2^{(0)}=(18.0\pm0.2)\times10^{-4}\,M_{\pi}^{-5}$, 
 $a_2^{(2)}=(2.2\pm0.2)\times10^{-4}\,M_{\pi}^{-5}$ and we have improved the 
value of $a_1$ by combining the independent determinations 
of this quantity from 
$F_\pi$ and with the Froissart--Gribov representation to get 
$a_1=(38.3\pm0.8)\times 10^{-3}\,M_{\pi}^{-3}$.} of ref.~13 for $a_1$ and the $a_2^{(I)}$, 
we find 
$$\langle r^2_{{\rm S},\pi}\rangle=0.83\pm0.17\;{\rm fm}^2,$$
compatible with our best value (3.14). 
If we had used the relation between $\langle r^2_{{\rm S},\pi}\rangle$ and the S wave phase shifts, 
$$1+\tfrac{1}{3}M^2_\pi\langle r^2_{{\rm S},\pi}\rangle=\dfrac{4\pi}{3}
\dfrac{f^2_\pi}{M^2_\pi}\Big\{2a_0^{(0)}-5a_0^{(2)}\Big\}
-\tfrac{41}{192}\,\dfrac{M^2_\pi}{\pi^2 f^2_\pi},
$$
 we would have obtained  somewhat larger numbers,
$$\langle r^2_{{\rm S},\pi}\rangle=1.26\pm0.26\;{\rm fm}^2,\quad 1.14\pm0.21\;{\rm fm}^2$$
depending on whether we use the scattering lengths themselves or the value for 
the combination $2a_0^{(0)}-5a_0^{(2)}$ from the dispersive integral in the Olsson sum rule 
[ref.~13(a), \equn{(4.3)}]. These values are a bit more than 
$1.5\,\sigma$ above (3.14): doubtlessly because the higher order corrections 
are larger for the  relation involving the S waves scattering lengths.

\vfill\eject
\brochuresection{Acknowledgments}

\noindent
I am grateful to J.~R.~Pel\'aez who triggered my interest in the subject by remarking the 
inconclusive status of existing evaluations of the scalar radii, as well as for 
a number of suggestions, in particular bringing to my attention 
the existence of the $\kappa$ enhancement, ref.~12.

\brochuresection{References}

\item{1 }{Gasser, J., and Leutwyler, H., {\sl Ann. Phys.} (N.Y.), {\bf 158}, 142  (1984). 
Second order calculations of both the pion and mixed form factors have appeared recently: 
Moussallam,~B., {\sl Eur. Phys. J.}, {\bf C14}, 111 (2000), 
Ananathanarayan,~B., Buttiker,~P., and Moussallam,~B., {\sl Eur. Phys. J.}, {\bf C22}, 133 (2000),
Jamin, M, Oller, J.~A., and Pich, A., {\sl Nucl. Phys.},{\bf B266}, 279 (2002), 
and  
Bijnens,~J., and Dhonte,~P., LU~TP~03-32 (2003), hep-ph/0307044. 
They do not add anything essential in connection with what interests us here.}
\item{2 }{Gasser, J., and Leutwyler, H., {\sl Nucl. Phys.}, {\bf B250}, 517 (1985).}
\item{3 }{Particle Data Tables: D. E. Groom et al., {\sl Eur. Phys. J.}, {\bf C15}, 1 (2000).}
\item{4 }{Artemov, V. M., et al., {\sl Phys. Atom. Nucl.}, {\bf 60}, 2023 (1997); 
Whitman, R., et al., {\sl Phys. Rev.}, {\bf D21}, 652 (1980); 
Heintze, J., et al., {\sl Phys. Letters}, {\bf 70B}, 482 (1977); 
Braun, et al., {\sl Nucl. Phys.}, {\bf B89}, 210 (1975).}
\item{5 }{Donoghue, J. F., Gasser, J.,  and Leutwyler, H., {\sl Nucl. Phys.}, {\bf B343}, 341 (1990).}
\item{6 }{Colangelo, G., Gasser, J.,  and Leutwyler, H.,
 {\sl Nucl. Phys.}, {\bf B603},  125 (2001).}
\item{7 }{Atkinson, D., Mahoux, G., and Yndur\'ain, F. J., {\sl Nucl. Phys.}, {\bf B54}, 263 
 (1973); 
{\bf B98}, 521 (1975). 
This ambiguity is far from being academic. For example, 
experimental phase shift analyses give a P wave in $\pi\pi$ scattering 
both essentially elastic, and clearly smaller than $\pi$ for $s\lsim4\,{\gev}^2$. 
This two properties, however, are unlikely to hold simultaneously: this energy 
should be  already asymptotic, and then 
the theorem (2.5) implies that one should have $\delta_1(t)=\delta(t)>\pi$.}
\item{8 }{Yndur\'ain, F. J.,
 {\sl  Nucl. Phys.}, {\bf B88}, 316 (1975); Aguilar-Ben\'{\i}tez,~M., et al., 
{\sl Nucl. Phys.}, {\bf B140}, 73 (1978).}
\item{9 }{Brodsky, S. J., and Farrar, G., {\sl Phys. Rev. Lett.}, {\bf 31}, 1153  (1973);  
Farrar, G., and Jackson, D. R., {\sl Phys. Rev. Lett.}, {\bf 43}, 246 (1979).
See also 
Brodsky, S. J., and  Lepage, G. P., {\sl Phys. Rev.} {\bf D22}, 2157  (1980); 
Duncan, A., in {\sl Topical Questions in QCD}, {\sl Phys. Scripta}, {\bf23}, No. 5 (1981).}
\item{10 }{Muskhelishvili, N.~I., {\sl Singular
 Integral Equations}, Nordhoof, Groningen  (1958). A detailed discussion 
tailored for the present cases, may be found in 
Yndur\'ain, F. J.,  ``Low energy pion interactions", FTUAM~02-28 (hep-ph/0212282).}
\item{11 }{de Troc\'oniz, J. F., and Yndur\'ain, F. J., {\sl Phys. Rev.},  {\bf D65}, 093001
 (2002).} 
\item{12 }{Theory: Oller, J.~A., Oset, E., and Pel\'aez, J. R., {\sl Phys. Rev.}, {\bf D59}, 
074001 (1999) and 
G\'omez-Nicola,~A. and Pel\'aez,~J.~R.,  {\sl Phys. Rev.}, {\bf D65}, 
054009 (2002).
 Experiment: Aitala, E.~M., et al., {\sl Phys. Rev. Lett.}, {\bf 89}, 12801 (2002). 
We take the width  and mass of the $\kappa$ from this last reference. 
Although this resonance has not made it to the PDT, it appears very 
likely that (bona fides resonance or not) 
the S$\tfrac{1}{2}$ phase for  $K\pi$ 
scattering is strongly enhanced at an energy a little below 
1~\gev.}
\item{13 }{(a) Pel\'aez, J. R., and Yndur\'ain, F. J., hep-ph/0304067 (to be 
published in Phys. Rev. D). See also 
(b) Pel\'aez, J. R., and Yndur\'ain, F. J., FTUAM 03-14, 2003 
(especially Appendix~A for phase shifts).}
\item{14 }{(a) Protopopescu, S. D., et al., {\sl Phys Rev.} {\bf D7}, 1279 (1973); 
(b) Hyams, B., et al., {\sl Nucl. Phys.} {\bf B64}, 134 (1973); 
Grayer, G., et al.,  {\sl Nucl. Phys.}  {\bf B75}, 189 (1974). See also the analysis of the 
same experimental data in
Estabrooks, P., and Martin, A. D., {\sl Nucl. Physics}, {\bf B79}, 301  (1974).}
\item{15 }{(BCT) Bijnens, J., Colangelo, G., and Talavera, G., 
{\sl JHEP}, {\bf 9805}, 014 (1998);
(ABT) Amor\'os, G.,  Bijnens, J., and Talavera, G., {\sl Nucl. Phys.}, {\bf B585}, 293 (2000).}
\item{16 }{Descotes, S., Fuchs, N. H.,  Girlanda, L., and   Stern, J., {\sl Eur. Phys. J. C}, 
{\bf 15}, 469 (2002).}
\item{17 }{Gasser, J., and Mei\ss ner,~U.~G., {\sl Nucl.~Phys.}, {\bf B357}, 90 (1991); 
Colangelo,~G., Finkelmeir,~M., and Urech,~R. {\sl Phys.~Rev.}, {\bf D54}, 4403 (1996); 
Frink,~M., Kubis,~B., and Mei\ss ner,~U.~G., {\sl Eur.~Phys.~J.}, {\bf C25}, 259 (2002).}

\bye